      [1999/12/01 v1.4c Il Nuovo Cimento]
\documentclass{cimento}

\usepackage{graphicx}

\newcommand{\rmd}{{\rm d}}

\newcommand{\pT}{p_{\rm T}}
\newcommand{\pTt}{p_{\rm Tt}}
\newcommand{\pTa}{p_{\rm Ta}}
\newcommand{\pTpair}{p_{\rm T,pair}}
\newcommand{\pout}{p_{\rm out}}
\newcommand{\vecpTt}{\vec{p}_{\rm Tt}}
\newcommand{\vecpTa}{\vec{p}_{\rm Ta}}
\newcommand{\hatpTt}{\hat{p}_{\rm Tt}}
\newcommand{\hatpTa}{\hat{p}_{\rm Ta}}

\newcommand{\jT}{j_{\rm T}}
\newcommand{\kT}{k_{\rm T}}

\title{First Jet and High $\pT$ Measuerements with the ALICE
  Experiment at the LHC}

\author{C.~Klein-B{\"o}sing\from{ins:a}\from{ins:b}\thanks{for the
    ALICE Collaboration}}
\instlist{\inst{ins:a}Institut f{\"u}r Kernphysik -  M{\"u}nster, Germany
\inst{ins:b}ExtreMe Matter Institute, GSI - Darmstadt, Germany}
\PACSes{\PACSit{13.85.}{t}
\PACSit{25.75.Dw}{}
}

\begin{document}

\maketitle

\begin{abstract}
The Large Hadron Collider at CERN currently provides p$+$p collisions
at center of mass energies of $\sqrt{s} = 7$~TeV, which allow to study high
  $\pT$ particle production and jet properties in a new energy
  regime. For a clear interpretation and the quantification of the medium
  influence in heavy-ion collisions on high $\pT$ observables a
  detailed understanding of these elementary reactions is
  essential. We present first results on the observation of jet-like
  properties with the ALICE experiment and discuss the performance of
  jet reconstruction in the first year of data taking.
\end{abstract}

\section{Introduction}

The ALICE experiment is the only dedicated heavy-ion experiment at the
Large Hadron Collider (LHC).  Its major goal is to study the creation of a
new state of strongly interacting matter, the Quark-Gluon Plasma
(QGP), which is expected to form when sufficiently high initial energy
densities are attained in collisions of heavy nuclei.

To study the full evolution of the produced medium the usage of so
called \emph{hard probes} is of particular interest. Parton-parton
scatterings with large momentum transfer $Q^2$ (\emph{hard})\cite{Berman:1971xz} occur in
the early stages of the reaction, in heavy-ion collisions well before
the formation of an equilibrated medium. While in p$+$p collisions
the partons evolve in the QCD-vacuum and fragment directly into jets
of observable hadrons, the evolution in heavy-ion collisions is
influenced by the presence of a medium with large density of color
charges. Thus, scattered partons can lose energy via medium induced
gluon radiation or elastic scattering, with the amount of energy loss
depending on the color charge and mass of the parton, the traversed
path length, and the medium-density \cite{Bjorken:1982tu,Gyu90a,Gyulassy:2003mc,Kovner:2003zj}.

One of the earliest predicted consequences has been the suppression of
hadrons with large transverse momentum ($\pT$) compared to the
expectation from scaled p+p reactions \cite{Bjorken:1982tu,Gyu90a}. This comparison is
usually done in a ratio of cross sections in p$+$p and $A+A$, the
nuclear modification factor $R_{AA}$:
\begin{equation}
\label{eq:raa}
R_{AA} = \frac{\rmd^2N_{AA}/ \rmd y \rmd\pT} {T_{AA} \cdot \rmd^2
\sigma_{\rm pp}/ \rmd y \rmd\pT},
\end{equation}
where $T_{AA}$ accounts for the increased number of nucleons in the
incoming $A$-nuclei and is related to the number of binary collisions
$N_{\rm coll}$ by $T_{AA} \approx N_{\rm coll}/\sigma_{\rm inel}^{\rm
  pp}$. In the absence of any medium effects and at sufficiently high
$\pT$, where hard scattering is the dominant source of particle
production, $R_{AA}$ should be unity, any deviation from unity
indicating the influence of the medium.

Indeed, already the first hadron measurements in Au$+$Au reactions at
RHIC showed a hadron yield suppressed up to a factor of five and a
suppression of jet-like correlation in central
collisions \cite{Adcox:2001jp,Adler:2002tq,Adler:2003qi,Ada03b}. This
suppression can be explained by the energy loss of hard scattered
partons via induced gluon bremsstrahlung in a medium with high color
density, which is also supported by the absence of suppression in the
yield of direct photons \cite{Adler:2005ig}.

Up to now, most measurements of the medium modification of hard scattered partons
are based on single particle measurements, which have been
used to deduce jet-like properties such as back-to-back correlations,
spectral shape/yield and their changes in heavy-ion collisions. This
has the disadvantage that the population of each single particle $\pT$-bin is highly biased towards a hard fragmentation. The bias can be
largely reduced by reconstruction of full jets, which also allows for a
detailed investigation of the expected modification of the jet
structure \cite{Salgado:2002ws} by
comparing momentum distributions of particles in jets created in $A +
A$ and p$+$p reactions. However, the measurement of jets in heavy-ion
collision is hindered by the presence of a large background and its
fluctuation. It recently succeeded for the first time at RHIC
energies of $\sqrt{s_{\rm NN}} = 200$~GeV (see
e.g. \cite{Salur:2009vz}), but the potential of full jet
reconstruction in heavy-ion collisions for a more quantitative
understanding of the medium properties will only be exploited at LHC
energies where the increase in jet cross section allows for a better separation
of the jet signal from the background in a larger kinematical range.

However, for a clear interpretation of all high $p_T$ observables in
heavy-ion collision a detailed understanding of p$+$p reactions is
essential. The comparison to predictions for the new energy regime
reached at the LHC in p+p collisions at $\sqrt{s} = 7$~TeV, provides
an important test of our theoretical understanding of the underlying
processes in these elementary collisions. The ALICE detector with its
excellent tracking and PID capabilities provides a versatile tool for the
studies of jet structure and composition over a large dynamic range
from jet transverse momenta of about 100~GeV/$c$ during the first LHC run, down to
particle $p _T$ of 100 MeV/$c$ \cite{Aamodt:2008zz}.

\section{Trigger-Particle Correlations}

\begin{figure}[th]
\centerline{\includegraphics[width=\textwidth]{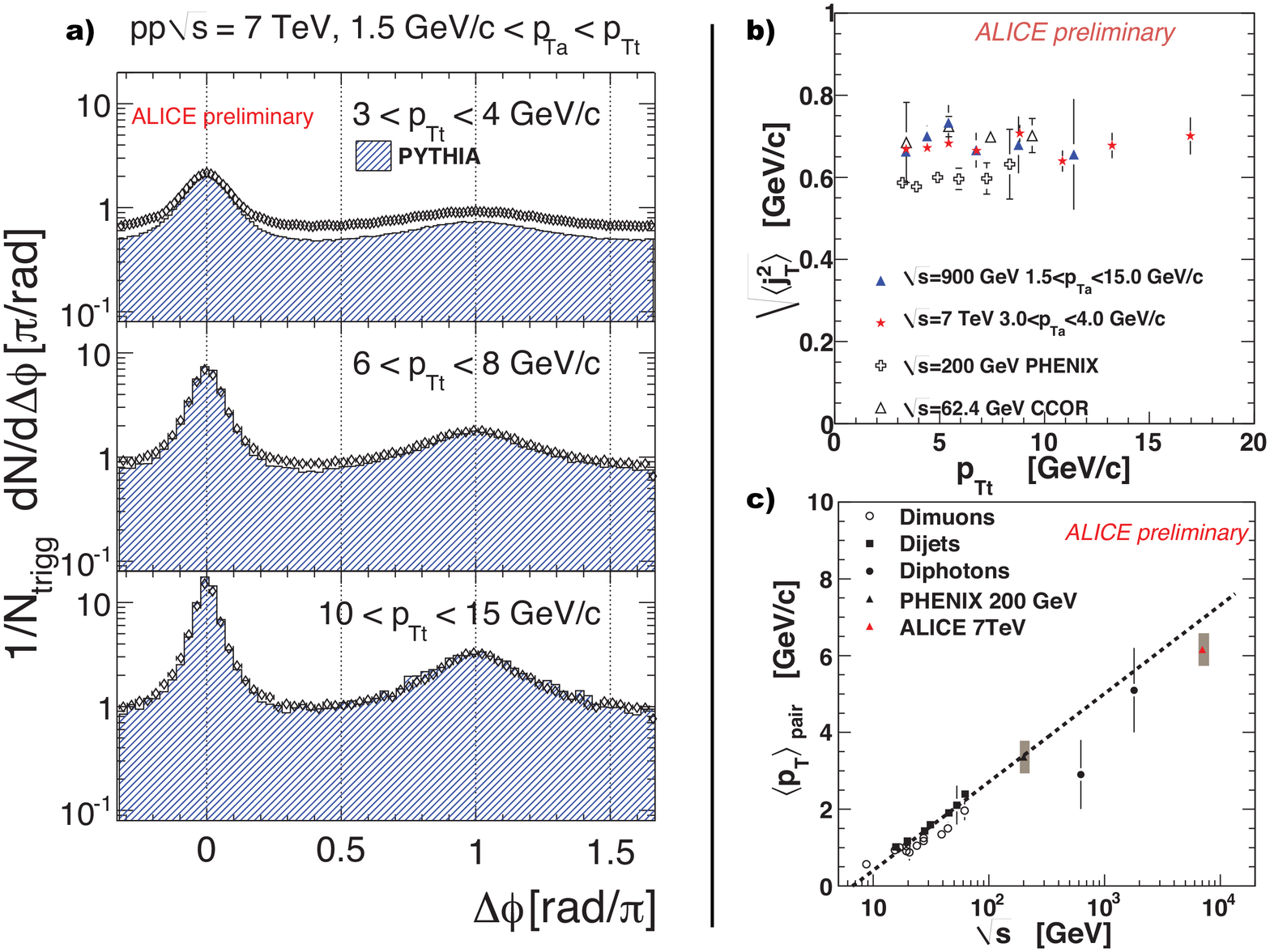}}
\vspace*{-15pt}
\caption{ 
\label{fig:CF_7TeV_jT_kT_abc}
a) Correlation function for three different ranges of trigger $\pT$ in
p$+$p collisions at $\sqrt{s_{NN}} = 7$~TeV 
compared to a PYTHIA simulation. b) $\sqrt{\langle \jT^2 \rangle }$ vs. the trigger $\pT$ measured with
the ALICE Experiment and compared to the values from CCOR and
PHENIX\protect\cite{Angelis:1980bs,Adler:2006sc}.  
c) $\sqrt{\langle \pTpair^2 \rangle }$  measured by ALICE at
7~TeV, compared to measurements at other $\sqrt{s}$.
}
\end{figure}
Particle correlations provide access to jet properties on an
inclusive basis and have been traditionally used in kinematical
regions where full jet reconstruction is difficult (i.e. lower center of
mass energy, low transverse momenta). In the large background
environment of heavy-ion collisions they provide the means to study
jet-like properties down to very low $\pT$ but at the cost of a
trigger bias towards hard fragmentation and jets with only little
energy loss.


Particle correlation functions have been measured by ALICE in p+p
collisions at $\sqrt{s} = 900$ and $7,000$~GeV. An example is shown in
Fig.~\ref{fig:CF_7TeV_jT_kT_abc}\,a), where the average correlation with respect
to a trigger particle in a given trigger $\pTt$ is shown. The
associated particles are chosen in a $\pT$ range from $1.5$~GeV$/c < \pTa < \pTt$. The expected back-to-back structure is
clearly seen down to the lowest $\pTt$: a peak in direction of the
trigger particle ($\Delta\phi = 0$, \emph{near side}) and a peak at
$\Delta\phi = \pi$ (\emph{away side}).

The width of the distribution on the near side is only given by the
non-perturbative fragmentation process, which leads to a transverse
momentum component ($\jT$) of the produced hadrons with respect to
the direction of the original parton. Assuming a two dimensional
Gaussian distribution of the $x$ and $y$ components of the $\jT$ vector,
one can extract its mean value directly from the near-side width
$\sigma_{\rm N}$ \cite{Adler:2006sc}:
\begin{equation}
\sqrt{\langle j_T^2 \rangle} =  \sqrt{2\langle j_{T,y}^2 \rangle}\approx \sqrt{2}
\frac{p_{Tt}p_{Ta}}{\sqrt{p_{Tt}^2 + p_{Ta}^2}}  \sigma_N .
\end{equation}
The results for different $\pTt$ and the two collision energies
measured by ALICE are shown in Fig.~\ref{fig:CF_7TeV_jT_kT_abc}\,b)
together with measurements from lower center of mass energies. Overall
no change with either the trigger $\pT$ nor the center of mass energy
is observed.

Due to the fact that the trigger particle direction fixes the
near-side 
axis all effects of interjet-correlations are seen on the
away-side. In particular deviations from the exact momentum balance of
the outgoing partons lead to additional broadening of the away-side
peak. It is expressed in terms of a net transverse momentum of the
parton pair $\langle \pT^2 \rangle_{\rm pair} = 2 \langle \kT^2
\rangle$. Here, $\langle \kT \rangle$ denotes the effective magnitude
of the apparent transverse momentum of each colliding parton caused
mainly by a combination of the intrinsic transverse momentum of the
incoming partons in the nucleons with initial and final state
radiation.  The momentum component of the away-side particle
($\vecpTa$) with respect to the trigger particle ($\vecpTt$) in the
transverse plane is called $\pout$. Its average value is related to
the away-side width but can also be measured directly and is used to
extract the magnitude of the momentum imbalance following \cite{Adler:2006sc}:
\begin{equation}
\frac{\langle z_{\rm t} \rangle}{\langle \hat{x}_{\rm h} \rangle}\sqrt{\langle {\kT^2} \rangle}
 = \frac{1}{x_{\rm h}} \sqrt{\langle {\pout^2} \rangle - \langle
   j_{\rm Ty}^2 \rangle (1 + x_{\rm h}^2)},
\end{equation}
where $x_{\rm h} = \pTa/\pTt$ and all values on the right hand side of
the equation are measured. The mean momentum fraction from
the original parton that is carried away by the trigger hadron
$\langle z_{\rm t} \rangle$ can be calculated based only on the shape
of the fragmentation function. $\langle \hat{x}_{\rm h} \rangle =
\hatpTa/\hatpTt$ denotes the imbalance between near and away-side
momentum already at parton level. Since it depends on the magnitude of 
$\kT$ it has to be determined iteratively. This procedure results in
the average momentum imbalance of the parton pair at the highest
$\sqrt{s}$ measured so far, it is shown in Fig.~\ref{fig:CF_7TeV_jT_kT_abc}\,c)
and confirms the logarithmic increase of the net transverse momentum
of the parton pair with the center of mass energy
in nucleon-nucleon collisions.

\section{Fully Reconstructed Jets}

Jets in ALICE are reconstructed using a variety of different
algorithms: e.g. sequential recombination algorithms
(anti-$\kT$ and $\kT$), simple cone algorithms as well as the
modern seedless and infrared safe cone (SISCone) algorithm. The
variety is motivated by the different sensitivity to the background
and different background subtraction schemes enabled by the various
algorithms \cite{Collaboration:2010ze,Alessandro:2006yt}.
\begin{figure}[th]
\centerline{\includegraphics[width=8cm]{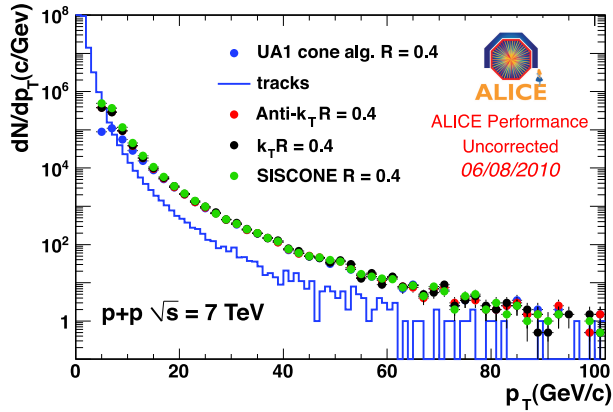}}
\vspace*{-5pt}
\caption{
\label{fig:2010-08-06_AlicePerf_jets_7TeV_v3}
The raw number of jets reconstructed in $|\eta| < 0.5$ with different
algorithms and $R = 0.4$, using 128M minimum bias p$+$p collisions at
$\sqrt{s} = 7$~TeV. As comparison the input raw track momentum
spectrum ($|\eta| < 0.9$) is also shown. Neither the jet energy-scale
has been corrected, nor any other acceptance and efficiency
corrections have been applied. }
\end{figure}

The definitive procedure for measuring jets in ALICE will be via
the combination of the central tracking system and
the calorimetric information from the EMCAL \cite{Collaboration:2010ze}.
Since the EMCAL represents an upgrade to the original ALICE setup, it
has not been fully installed for the first period of data taking and
currently only covers a limited acceptance. 
Thus the jet reconstruction in ALICE at present is based on the information from reconstructed tracks, i.e. only charged
particles. When comparing these reconstructed jets to full particle
jets the momentum scale is shifted to lower values by roughly a
factor of 0.6 due to the missing neutral energy and the jet resolution
is dominated by charged-to-neutral fluctuations in addition to the
instrumental momentum resolution. The raw jet spectrum (not corrected
for the aforementioned effects) from reconstructed tracks is shown in
Fig.~\ref{fig:2010-08-06_AlicePerf_jets_7TeV_v3} for different
algorithms with a resolution parameter or cone size of $R = 0.4$ within $|\eta| <
0.5$, the input tracks are taken in the acceptance of $|\eta| < 0.9$.
The results from all jet finders agree well in the region above $\pT =
20$~GeV/$c$, where the effects of seed particles (UA1 cone algorithm)
and split-merge algorithm (SISCone) become negligible, illustrating
that in p$+$p collisions all jet definitions lead to a uniform
picture. This will be tested again in Pb$+$Pb collisions, where the
different susceptibility to background is more important.


The final measurement of the jet spectrum with charged particles will employ an unfolding
method to fully account for the jet response of the ALICE
detector. The corrections will be done up to the level of charged
particles accounting mainly for instrumental resolution, acceptance
and efficiency, which can be compared to the same measurement in
Pb$+$Pb. The correction up to the level of jets from all particles
introduces a larger uncertainty due to the magnitude of the
charged-to-neutral fluctuations, but it will allow to compare more
 directly to pQCD calculations and jet measurements by other
experiments. The limitation in the response due to the missing neutral energy will be
overcome with the addition of the EMCAL information after its
completion to full acceptance.

\section{Jet Finding in Pb+Pb}

The first essential measurement in Pb$+$Pb collisions at $\sqrt{s_{\rm
    NN}} = 2.75$~TeV, which will be recorded in November 2010, is the
determination of the overall multiplicity and thereby the level of
background for the jet reconstruction and its fluctuation. In
simulations for central Pb$+$Pb collisions at $\sqrt{s_{\rm NN}} =
5.5$~TeV the total amount of background energy is about 200~GeV in an
area corresponding to $R \approx 0.4$, this can be determined and
subtracted on a event-by-event basis. The fluctuations in the
background within  one event have been estimated to have a width of $B_i^A
=12$~GeV when using the same area as above \cite{Collaboration:2010ze}. These can
be corrected for in the inclusive jet measurement via an unfolding
procedure, enabling the comparison of the jet spectra in p$+$p and
Pb$+$Pb.
\begin{figure}[h]
\centerline{\includegraphics[width=0.90\textwidth]{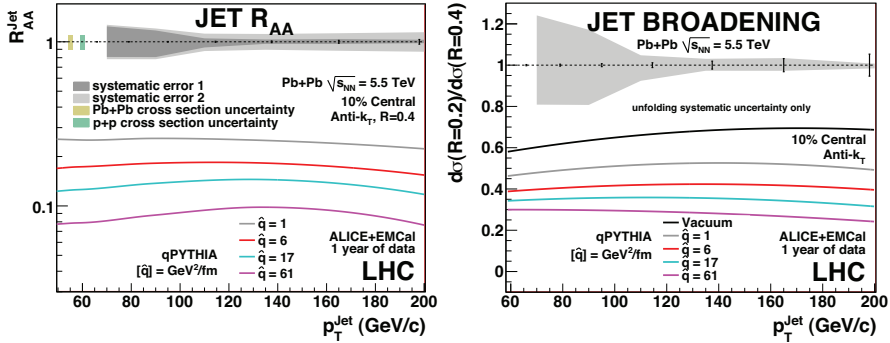}}
\caption{
\label{fig:PPR_JetRAABroad}
Expected performance of ALICE jet measurements in one year of nominal
Pb$+$Pb data taking,  jets measured with the anti-$\kT$ algorithm
and $R = 0.4$.  Left: Jet $R_{\rm AA}^{\rm jet}$
simulated for different $\hat{q}$ with Q-PYTHIA at $\sqrt{s_{NN}} =
5.5$~TeV.  Right: Ratio of inclusive differential jet cross sections
at different $R$ simulated for various $\hat{q}$ with
Q-PYTHIA \protect\cite{Collaboration:2010ze}.}
\end{figure}
The expected magnitude and precision of the nuclear modification
factor for jets in one nominal year of data taking and including the
EMCAL, is shown in Fig.~\ref{fig:PPR_JetRAABroad} for different
values of the medium transport coefficient
$\hat{q}$ \cite{Collaboration:2010ze}. It is clearly seen that jet
measurements in ALICE will be sensitive to the change of $R_{\rm
  AA}^{\rm jet}$ with increasing energy loss. A more discriminative measure,
which will help to verify the effect of jet broadening is the ratio of
the jet yields obtained with different resolution parameters $R$. It
is also shown in Fig.~\ref{fig:PPR_JetRAABroad} and illustrates how
the energy within the jet is redistributed to larger distances from
the jet axis, in this particular model due to modified splitting
functions for the parton shower evolution in the medium
\cite{Armesto:2009fj}. These measurements will be complemented by a
detailed comparison of momentum distributions within jets.

\section{Conclusions}
We have presented the first results on particle correlations
measured by the ALICE experiment in p$+$p collisions at the
LHC, they give access to jet properties in a kinematical region which
is difficult to access with full jet finding and provide the first
measurement of the momentum imbalance of parton pairs in this energy regime. The detailed
characterisation of reconstructed jets in p$+$p events at 7~TeV is currently
going on and we await the first reconstruction of jets in heavy-ion
collisions at the LHC in the fall of 2010.

\section*{Acknowledgements}

This work was supported by the Alliance Program of the
Helmholtz Association (HA216/EMMI).

\end{document}